\documentclass[11pt]{article}
\usepackage{moriond,epsfig,amsfonts,latexsym}

\def\IZ{\mathbb{Z}}

\def\be{\begin{equation}}
\def\ee{\end{equation}}
\def\bea{\begin{eqnarray}}
\def\eea{\end{eqnarray}}

\begin{document}
\vspace*{4cm}
\title{STRINGS, GRAVITY AND PARTICLE PHYSICS}

\author{ A. SAGNOTTI${}^1$ AND A. SEVRIN${}^2$ \vskip 12pt}

\address{${}^1$Dipartimento di Fisica\\Universit\`a di Roma ``Tor Vergata''\\
INFN- Sezione di Roma ``Tor Vergata''\\ Via della Ricerca Scientifica,1 \ 
00133 Roma, ITALY \vskip 12pt
${}^2$ Theoretische Natuurkunde\\ Vrije Universiteit Brussel\\
Pleinlaan 2, B-1050 Brussels, BELGIUM}

\maketitle\abstracts{
This contribution, aimed mostly at experimental 
particle physicists, reviews some of the main
ideas and results of String Theory in a non-technical language. It originates
from the talks presented by the authors at the Electro-Weak session of the 
2002 Moriond Meeting, here merged in an attempt to provide a more complete
and concise view of the subject.
}
\section{Introduction}

One of the main achievements of Physics is certainly the reduction of
all forces in Nature, no matter how diverse they might appear at first sight, 
to four fundamental types: gravitational, electromagnetic, weak and
strong. The last three, in particular, are nicely described by the 
Standard Model, a Yang-Mills gauge theory where the gauge group ${\rm SU(2)}_L
\times {\rm U(1)}_Y\times {\rm SU(3)}_{QCD}$ is 
spontaneously broken to ${\rm U(1)}_{em}\times {\rm SU(3)}_{QCD}$. 
A gauge theory is a generalization of Maxwell's theory of Electromagnetism 
whose matrix-valued potentials satisfy non-linear field equations even
in the absence of matter, and the
corresponding gauge bosons are the quanta associated to their wave modes. 
For instance, the $W$ and $Z$ bosons, quanta of the corresponding 
$W_\mu$ and $Z_\mu$ gauge fields, are charged under one or more of the previous
gauge groups, and are thus mutually interacting, an important
feature well reflected by their
non-linear field equations. The other key ingredient of the Standard
Model, the spontaneous breaking of ${\rm SU(2)}_L
\times {\rm U(1)}_Y$ to ${\rm U(1)}_{em}$, is a sort of Meissner 
effect for the whole of 
space time, that is held responsible for screening the weak force down 
to very short distances. It relies on a universal low-energy description
of the phenomenon in terms of scalar modes, and therefore the search for
the residual Higgs boson (or, better, Brout-Englert-Higgs or BEH boson)
is perhaps the key effort in experimental
Particle Physics today. Whereas the resulting dynamics is very 
complicated, the 
Standard Model is {\it renormalizable},
and this feature allows reliable and consistent perturbative analyses
of a number of quantities
of direct interest for Particle Physics. These are by now tested by
very precise experiments, as we have heard in several Moriond talks, and
therefore, leaving aside the BEH boson that is yet to
be discovered, a main problem today is ironically
the very good agreement between the current experiments and the Standard Model,
with the consequent lack of clear signals for new Physics
in this domain. 

Despite the many successes of this framework, a number of aesthetic and 
conceptual issues have long puzzled 
the theoretical physics community: in many respects the
Standard Model does not have a compelling structure, while
gravity can not be incorporated in a satisfactory fashion.
In fact, gravity differs in crucial respects from the other 
fundamental forces, since it is very weak and 
plays no role in Atomic and Nuclear Physics: for instance,
the Newtonian attraction in a hydrogen atom is lower than the corresponding 
Coulomb force by an astonishing factor, 42 orders of magnitude. Moreover, the 
huge ratio
between Fermi's constant $G_F$ and Newton's constant $G_N$, that determine
the strength of the weak and gravitational interactions at low energies,
$G_F/G_N \sim 10^{35}\hbar^2 c^{-2} $, poses by itself a big puzzle, usually called the 
{\it hierarchy problem}: it is unnatural to have such a large
number in a fundamental theory, and in addition virtual quantum effects in the 
vacuum mixing the different interactions would generally make such a choice
very unstable. {\it Supersymmetry}, an elegant symmetry between boson 
and fermion modes introduced in this context by J. Wess and B. Zumino in 
the early seventies, can alleviate the problem by stabilizing the hierarchy, 
but does not eliminate the need for such unnatural constants.
It also predicts the existence
of Fermi and Bose particles degenerate in mass, and therefore it
can not be an exact feature of our low-energy world, while
attaining a fully satisfactory picture of supersymmetry breaking is a 
major challenge in present attempts.

In sharp contrast with the other three fundamental forces, Newtonian 
gravity is purely attractive, so that despite its weakness in the 
microscopic realm it dominates the large-scale dynamics
of our universe. General Relativity encodes these infrared 
properties in a very elegant way and,
taken at face value as a quantum theory, it would associate to the 
gravitational interaction an additional
fundamental carrier, the graviton, that would be on the same
footing with the photon, the gluons and the intermediate $W$ and $Z$
bosons responsible
for the weak interaction. The graviton would be a massless spin-two 
particle, and the common tenet is that
its classical Hertzian waves have escaped a direct detection 
for a few decades only
due to their feeble interactions with matter. Differently from the Standard
Model interactions, however,
General Relativity is not renormalizable, essentially because the 
gravitational interaction between point-like carriers  that, as we shall see in
more detail at the end of Section 2, is measured by the 
effective coupling
\begin{equation}
\alpha_N(E)\sim G_N E^2/\hbar c^5 \ , \label{gravalpha}
\end{equation}
grows rapidly with
energy, becoming strong at the Planck scale $
E_{Pl} \approx 10^{19} GeV$, defined so that
$\alpha_N(E_{Pl})\approx 1$. This scale, widely
beyond our means of investigation if not of imagination itself, 
is in principle explored by virtual quantum processes,
and as a result unpleasant divergences arise in the quantization of
General Relativity, that in modern terms seems to provide at most
an effective description of gravity at energies well below
the Planck scale. This is the {\it ultraviolet problem} of Einstein gravity,
and this state of affairs is not foreign. Rather, it is somewhat 
reminiscent of how the Fermi theory describes 
the weak interactions well below the mass scale of the intermediate bosons,
$E_W \approx 100 \ GeV$, where the effective Fermi coupling 
$\alpha_F(E) \sim G_F E^2/\hbar^3c^3$
becomes of order one. It is important to keep in mind that this analogy, 
partial as it may be, lies at the heart of 
the proposed link between String Theory and the fundamental interactions.

String Theory provides a rich framework for connecting gravity
to the other forces, and indeed it does so in a way that 
has the flavor of the modifications introduced by the Standard Model 
in the Fermi interaction: at the Planck scale new 
states appear, in this case actually an infinity of them,
that result in an effective weakening of the gravitational force. This
solves the ultraviolet problem of four-dimensional gravity, but 
the resulting picture, still far from complete, raises a number of 
puzzling questions that still lack a proper answer and
are thus actively investigated by many groups. One long-appreciated surprise,
of crucial importance for the ensuing discussion, is that String Theory, in
its more popular, or more tractable, supersymmetric version, requires that 
our space time include {\it six additional dimensions}. Despite the clear
aesthetic appeal of this framework, however, 
let us stress that, in dealing with matters that 
could be so far beyond the currently accessible scales,
it is fair and wise to avoid untimely conclusions, keeping also
an eye on other possibilities. These include a possible thinning
of the space-time degrees of freedom around the Planck scale, that
would solve the ultraviolet problem of gravity in a radically 
different fashion. For the Fermi
theory, this solution to its ultraviolet problem would assert the 
impossibility of processes entailing energies or momenta beyond the 
weak scale. While this is clearly not the case for weak interactions, 
we have no fair way to exclude that something of this sort could 
actually take place at the Planck scale, on which we
have currently no experimental clues. This can
be regarded as one of the key points of the canonical approach to quantum
gravity, long pursued by a smaller community of experts in General Relativity.

\begin{figure}
\begin{center}
\psfig{figure=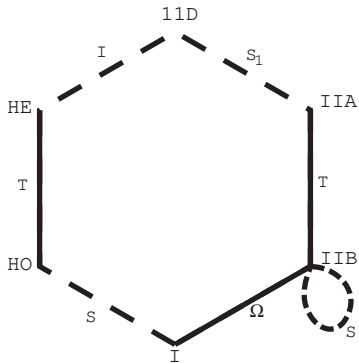}
\end{center}
\caption{The five ten-dimensional superstring theories are dual to one
another. The solid lines denote perturbative dualities, while the dashed ones 
indicate non-perturbative ones. At strong coupling, both the type 
$IIA$ and heterotic ${\rm E}_8\times {\rm E}_8$ strings develop an additional large
dimension,
a circle ($S_1$) and an interval ($I_1$) respectively. Therefore,
they are both described by an eleven dimensional theory, but this 
bears no direct relation to strings.
\label{fig:duality}}
\end{figure}

With this proviso, we can return to String Theory, the main theme of our 
discussion. Ideally, one should demand from it two things: some 
sort of 
uniqueness, in order to make such a radical departure from the Standard Model,
a four-dimensional Field Theory of point particles, more compelling, and some
definite path for connecting it to the low-energy world. The first goal has
been achieved to a large extent in the last decade, after the five
{\it supersymmetric} string models, usually called type IIA, type IIB, 
heterotic SO(32) (or, for brevity, HO), 
heterotic ${\rm E}_8 \times {\rm E}_8$ (or, for brevity, HE)
and type I, have been argued to
be equivalent as a result of surprising generalizations of 
the {\it electric-magnetic
duality} of Classical Electrodynamics. Some of these string dualities are 
nicely
suggested by perturbative String Theory, and in fact can also
connect other non-supersymmetric
ten-dimensional models to the five superstrings, while others
rest on the unique features of
ten-dimensional supergravity. Supergravity is an elegant extension of 
General Relativity, discovered in the mid seventies by S. Ferrara,
D.Z. Freedman and P. van Nieuwenhuizen, that describes the effective 
low-energy dynamics of the light superstring modes, 
where additional local supersymmetries require corresponding
gauge fields, the {\it gravitini}, and bring about, in general, other matter fields.
In ten dimensions, supergravity is {\it fully determined} by the type of
supersymmetry involved, (1,0), (1,1) or (2,0), 
where the numbers count the (left and right)
Majorana-Weyl ten-dimensional supercharges, and in the first case 
by the additional choice of a Yang-Mills gauge group, and this 
rigid structure allows one to make very
strong statements \footnote{This counting is often a 
source of confusion: in four dimensions a Weyl 
spinor has two complex, or four real,
components, while in ten dimensions the corresponding minimal Majorana-Weyl
spinor has sixteen real components, four times as many. Thus, the minimal
(1,0) ten-dimensional supersymmetry is as rich as ${\cal N}=4$ in
four dimensions, while a similar link holds between the (1,1) and (2,0) cases
and ${\cal N}=8$ in four dimensions.}. The end result is summarized in the
duality hexagon of figure \ref{fig:duality}, where the 
solid links rest on perturbative string arguments, while the 
dashed ones reflect non-perturbative features implied by ten-dimensional 
supergravity. The resulting
picture, provisionally termed ``M-Theory'', has nonetheless a puzzling feature: it links the
ten-dimensional superstrings to the eleven-dimensional 
Cremmer-Julia-Scherk (CJS) supergravity, that can be shown to bear no direct relation to strings!

An additional, vexing problem, is that
the reduction from ten dimensions 
to our four-dimensional space time entails
a deep lack of predictivity for the low-energy parameters, that depend on the
{\it size and shape} of the extra dimensions. This fact reflects
the absence of a global minimum principle for gravity, similar to those that
determine the ground states of a magnet in a weak external field below
its Curie temperature or the spontaneous
breaking of the electro-weak symmetry in the Standard Model, and represents a
stumbling block in all current approaches that aim at deriving 
our low-energy parameters from String Theory. 
It has long been hoped that a better understanding of string dynamics would
help bypassing this difficulty, but to date no concrete progress has been
made on this crucial issue.
Thus, ironically, for what we currently understand, String Theory appears 
to provide a unique answer to the problem of including gravity in
the Standard Model, but the four-dimensional remnants 
of this uniqueness are at least classes 
of theories. 
Supersymmetry has again a crucial effect on this problem, since it
basically {\it stabilizes} the internal geometry, much along the lines of
what we have seen for the hierarchy between the electro-weak and Planck
scales, but as a result the sizes and shapes (moduli) of
the extra dimensions are apparently {\it arbitrary}. 
This is the {\it moduli problem} of supersymmetric vacua, a problem indeed,
since the resulting low-energy parameters generally depend on the moduli.
On the other hand,
the breaking of supersymmetry, a necessary ingredient to recover the Standard 
Model at low energies if we are to describe Fermi and Bose fields of different
masses, tends to destabilize the background space time.
The end result is that, to date, although we know a number of
scenarios to break supersymmetry within String Theory, that we
shall briefly review in Section 5, we have little or
no control on the resulting space times once quantum
fluctuations are taken into account. 

The following sections are devoted
to some key issues raised by the extension from
the Standard Model to String Theory, in an attempt
to bring some of the main themes of current research to the attention of the
interested reader, while  using as starting
points basic notions of Electrodynamics,  Gravitation and Quantum
Mechanics. Our main target 
will be our colleagues active in Experimental High-Energy Physics, in the 
spirit of the Moriond Meeting, a very beneficial confrontation 
between theorists and experimentalists active in Particle Physics today.
We hope that this short review will help convey to them the 
excitement and the difficulties faced by
the theorists active in this field.
\section{From particles to fields}

The basic tenet from which our discussion may well begin is that {\it 
all matter is apparently made of elementary particles}, while 
our main theme
will be to illustrate why this may not be the end of the story.
Particles exchange mutual forces, and 
the Coulomb force between a pair of static point-like charges $q_1$ and $q_2$,
\begin{equation}
\left| {\bf F}_C \right| \sim \frac{| q_1 \, q_2 |}{r^2} \ ,
\end{equation}
with an intensity proportional to their product 
and inversely proportional to the square of their mutual distance, 
displays a remarkable similarity with the
Newton force between a pair of static point-like masses $m_1$ and $m_2$,
\begin{equation}
\left| {\bf F}_N \right| \sim \frac{| G_N \, m_1 \, m_2 |}{r^2} \ .
\label{newtforc}
\end{equation}

Actually, it has long been found more convenient to think of these basic
forces in two
steps: some ``background'' charge or mass distribution affects the
surrounding space creating a {\it field} that, in its turn, can affect
other ``probe'' charges or masses, sufficiently small not to perturb 
the background significantly. 
In the first case, the classical dynamics is encoded in the
Maxwell equations, that relate the electric field ${\bf E}$ and the magnetic
field ${\bf B}$ to electric charges and currents, and as a result
both fields
satisfy in vacuum wave equations of the type
\begin{equation}
\frac{1}{c^2} \ \frac{\partial^2\phi}{\partial t^2}  - \nabla^2 \phi = 0 \ .
\label{wave}
\end{equation}
These entail retardation effects due to the finite speed $c$ with which
electromagnetic waves propagate, and, as first recognized by Lorentz and
Einstein, provide the route to Special Relativity.

With gravity, the situation is more complicated, since the resulting
field equations are highly non linear. According to Einstein's 
General Relativity,
the gravitational field is a distortion of the space-time geometry that
replaces the Minkowski metric $\eta_{\mu\nu}$ with
a generic metric tensor $g_{\mu\nu}$, used to compute the
distance between two nearby points as
\begin{equation}
ds^2 = g_{\mu\nu}(x)\, dx^\mu \, dx^\nu \ . \label{gmn}
\end{equation}
Material bodies follow {\it universally} curved trajectories that
reflect the distorted geometries, while
the metric $g_{\mu\nu}$ satisfies a set of non-linear wave-like
equations where the energy-momentum of matter appears as a source.
In fact, the non-linear nature of the resulting
dynamics reflects the fact that the gravitational field
carries energy, and is therefore bound to act as its own source.
These observations extend a familiar fact: in the local uniform
gravitational field ${\bf g}$ near the earth ground, Newtonian bodies 
fall according to
\begin{equation}
m_i \, {\bf a} = m_g \, {\bf g} \ ,
\end{equation}
and the equality of the inertial and gravitational masses $m_i$ and $m_g$
makes this motion {\it universal}. 
The resulting ``equivalence principle'' is
well reflected in the distorted space-time geometry, that has
inevitably a universal effect on test bodies.
The modification in (\ref{gmn}) can not be the whole
story, however, since a mere change of coordinates can do this
to some extent, a simple
example being provided by the transition to
spherical coordinates in three-dimensional
Euclidean space, that turns the standard Euclidean metric
\begin{equation}
ds^2 = dx^2 +dy^2 +dz^2 
\end{equation}
into
\begin{equation}
ds^2 = dr^2 + r^2 \, d\theta^2 + r^2 \, \sin^2 \theta \, d\phi^2 \ .
\end{equation}

This simple example reflects a basic ambiguity met when 
describing the gravitational
field via a metric tensor, introduced by the freedom available in the choice of
a coordinate system. Strange as it may seem, this is but another, if 
more complicated,
instance of the ambiguity met when describing the Maxwell equations in terms
of the potentials ${\bf A}$ and $\Phi$, defined via
\begin{equation}
{\bf B} = \nabla \times {\bf A} \ , \qquad \quad {\bf E} = - \nabla \Phi - \frac{1}{c} \, \frac{\partial {\bf A}}{\partial t} \ ,
\end{equation}
a familiar fact of Classical Electrodynamics.
This ambiguity, in the form of {\it gauge transformations} of parameter
$\Lambda$
\begin{equation}
{\bf A} \to {\bf A} + \nabla \Lambda \ , \qquad \quad \Phi \to \Phi - 
\frac{1}{c} \, \frac{\partial \Lambda}{\partial t} \ , \label{gauge.em}
\end{equation}
does not affect measurable quantities like ${\bf E}$ and ${\bf B}$. A
suitable combination of derivatives of $g_{\mu\nu}$, known as the 
Christoffel connection $\Gamma^\mu_{\nu\rho}$, is the proper gravitational 
analog of the electrodynamic potentials ${\bf A}$ and $\Phi$.
Notice the crucial difference: in gravity the potentials are derivatives of the
metric field, a fact that has very
important consequences, since it essentially determines Eq. \ref{gravalpha}. 
In a similar fashion, the gravitational counterparts of the
${\bf E}$ and ${\bf B}$ fields can be built from the Riemann curvature 
tensor $R^\mu{}_{\nu\rho\sigma}$, essentially a curl of 
the Christoffel connection $\Gamma^\mu_{\nu\rho}$,
that thus contains second derivatives of $g_{\mu\nu}$. Summarizing,
gravity manifests itself as a curvature of the space-time geometry, that 
falling bodies are bound to experience in their motion.

Notice that Eq. \ref{gauge.em} can also be 
cast in the equivalent form
\begin{equation}
{\bf A} \to {\bf A} + \frac{\hbar c}{i q} e^{-\frac{iq\Lambda}{\hbar c}} \, 
\nabla 
e^{\frac{iq\Lambda}{\hbar c}}  
\ , \qquad \quad \Phi \to \Phi - \frac{\hbar}{iq} \,  
e^{-\frac{iq\Lambda}{\hbar c}}
\frac{\partial\ }{ \partial t}\, e^{\frac{i q\Lambda}{\hbar c}} \ , \label{ngauge.em}
\end{equation}
a rewriting that has a profound meaning, since it
is telling us that in Electrodynamics the effective gauge parameter
is a pure phase,
\begin{equation}
\beta = e^{i\frac{q\Lambda}{\hbar c}} \ .
\end{equation}
Quantum Mechanics makes this interpretation quite compelling, as
can be seen by the following simple reasoning. In Classical Mechanics,
the effect of electric and magnetic fields on a particle of charge
$q$ is described
by the Lorentz force law,
\begin{equation}
{\bf F} \ = \ q\left( {\bf E} \ + \ \frac{1}{c}\, {\bf v} \times {\bf B}   \right) \ ,
\end{equation}
while Quantum Mechanics makes use of the Hamiltonian $H$ or of the Lagrangian
$L$, from which this force can obtained by differentiation. Thus, $H$ and
$L$ are naturally bound to involve the potentials, and so does the 
non-relativistic Schr\"odinger
equation
\begin{equation}
- \frac{\hbar^2}{2m} \, \left( \nabla - \frac{iq}{\hbar c} {\bf A} \right)^2
\, \psi \ + \ q \, \Phi \, \psi \ = \ i \hbar \, \frac{\partial \psi}{\partial t} \ ,
\label{schrod}
\end{equation}
that maintains its form after a gauge transformation only provided 
the wave function $\psi$ transforms as
\begin{equation}
\psi \, \to \, e^{\frac{iq}{\hbar c} \Lambda} \, \psi
\end{equation}
under the electromagnetic gauge transformation (\ref{ngauge.em}), thus
leaving the probability density $|\psi|^2$ unaffected. 
Notice that the electromagnetic fields can also be recovered from
commutators of the {\it covariant derivatives} in (\ref{schrod}):
for instance
\begin{equation}
\left[ \partial_i - \frac{iq}{\hbar c} {\bf A}_i, 
\partial_j - \frac{iq}{\hbar c} {\bf A}_j  \right] \ = \ - \frac{iq}{\hbar c} \,
\left(\partial_i {\bf A}_j - \partial_j {\bf A}_i \right) \ =\ 
- \, \frac{iq}{\hbar c} \, \varepsilon_{ijk} \, {\bf B}_k \ .
\label{commutator}
\end{equation}

If Special Relativity is combined with Quantum Mechanics, one is 
inevitably  led to a multi-particle description: quantum energy fluctuations 
$\Delta E \sim m c^2$ can
generally turn a particle of mass $m$ into another, and therefore one
can not forego the need for {\it a
theory of all particles of a given type}. Remarkably, 
the field concept is naturally
tailored to describe particles, for instance {\it all} the identical photons
in nature, and it does so in a relatively simple
fashion, via the theory of the harmonic oscillator. A wave equation emerges
in fact from the continuum limit of coupled harmonic
oscillators, a basic fact 
nicely reflected by the corresponding normal modes, as can be seen letting
\begin{equation}
\Phi({\bf x},t) \ = \ e^{i\,  {\bf k}\cdot {\bf x}} \, f(t)
\end{equation}
in Eq. \ref{wave}. Quantum Mechanics associates to the
resulting harmonic oscillators
\begin{equation}
\frac{d^2 f}{dt^2} \ + \ c^2 \, {\bf k}^2 \, f \ = \ 0 \, ,
\end{equation}
{\it equally spaced} spectra of excitations, that represent
{\it identical} particles, each characterized by a
momentum ${\bf p} = \hbar {\bf k}$, the 
{\it photons} in the present example. The allowed energies are
\begin{equation}
E_n({\bf k}) \ = \ \hbar \,c \,|{\bf k}| \, \left(n \, +\, \frac{1}{2} \right) \quad (n=0,1,...) 
\ ,
\end{equation}
and the equally spaced spectra allow an identification
of the $n$-th excited state with a collection of $n$ photons. Notice the
emergence of the zero-point energy $\frac{1}{2} \hbar c \, |\bf{k}|$, a 
reflection of the uncertainty principle
to which we shall return in the following. Let us add
that a similar reasoning for fermions would differ in two respects.
First, the Pauli principle would only allow $n=0,1$ for each ${\bf k}$,
while for the general case of massive fermions with momentum ${\bf p}=\hbar \,
{\bf k}$
the allowed energies would be in general 
\begin{equation}
E \ = \ \sqrt{c^2 {\bf p}^2 + m^2 c^4 }\, \left( n - \frac{1}{2} \right)  \quad (n=0,1) \ .
\end{equation}
Notice the {\it negative} zero-point energy, to be compared with the
{\it positive} zero-point energy for bosons. Incidentally, equal numbers 
of boson and fermion types degenerate in mass would result in an exactly
vanishing zero point energy, a situation realized in models with
{\it supersymmetry}. 

This brings us naturally to a brief discussion of the 
{\it cosmological constant
problem}, a wide mismatch between macroscopic and
microscopic estimates of the vacuum energy density in our universe. Notice 
that, in the presence of gravity, an additive contribution to the vacuum energy
has sizable effects: energy, just like mass, gives rise to gravitation, and
as a result a vacuum energy appears to endow the universe with 
a corresponding average curvature. 
Macroscopically, one has a time scale 
$t_H \sim 1/H \sim 10^{17} \, s$, where the Hubble constant $H$ 
characterizes the expansion rate of our universe, and a simple dimensional
argument associates to it an energy density 
$\rho_M \sim \frac{H^2 \; c^2}{G_N}$. One can attempt a theoretical 
estimate of this quantity, following Ya.B.~Zel'dovic, 
taking into account the zero-point energies
of the quantum fields that describe the types of particles present in
nature. A quantum field, however, even allowing no modes with 
wavelengths below the 
Planck length $\ell_{Pl}= \hbar c/E_{Pl}\approx 10^{-33}\, cm$, 
the Compton wavelength associated to the Planck scale, where 
as we have seen
gravity becomes strong, would naturally contribute via its zero-point 
fluctuations a Planck energy per
Planck volume, or $\rho_m \sim E_{Pl}^4/(\hbar c)^3$.
Using Eq. \ref{gravalpha} to relate $G_N$ to $E_{Pl}$, the ratio between
the theoretical estimate of the vacuum energy density and its 
actual macroscopic value is then
\begin{equation}
\frac{\rho_M}{\rho_m} \sim \left(  \frac{\hbar \; H}{E_{Pl}} \right)^2 \approx 
\ 
10^{-120} \ . \label{cosmconstprobl}
\end{equation}
This is perhaps the most embarrassing failure of contemporary physics, and
to many theorists it has the flavor of the  
blackbody problem,
where a similar mismatch led eventually to the
formulation of Quantum Mechanics.
In a supersymmetric world the
complete microscopic estimate would give a vanishing result since,
as we have seen, fermions and bosons give opposite contributions
to the vacuum energy. Still, with supersymmetry broken at a scale $E_s$
in order to allow for realistic mass differences $\delta M \sim E_s/c^2$
between bosons and fermions, 
one would essentially recover the previous estimate, but for the replacement
of $E_{Pl}$ with the supersymmetry breaking scale $E_s$, so that, say, 
with $E_s \sim 1 \ TeV$, the ratio in (\ref{cosmconstprobl}) would become
about $10^{-88}$, with an improvement of about 30 orders of magnitude. 
These naive considerations should suffice to motivate the current interest in
the search for realistic supersymmetric extensions of the
Standard Model with the lowest scale of supersymmetry breaking compatible
with current experiments where, accounting also for the contribution of 
gravity that here we
ignored for the sake of simplicity, more sophisticated
cancellations can allow to reduce the bound much 
further. We should stress, however, that no
widely accepted proposal exists today, with or without supersymmetry or
strings, to resolve this clash between Theoretical Physics and the observed 
large-scale structure of our universe.

We have thus reviewed
how all {\it identical} particles of a given type 
can be associated to the normal modes of a single field. While these
are determined by the 
{\it linear} terms in the field equations, 
the corresponding non-linear terms mediate transformations of
one particle species into others. This ``micro-Chemistry'', the object
of Particle Physics experiments, is regulated by
conservation laws, and in fact the basic reaction mechanisms
in the Standard Model
are induced by proper generalizations of the electromagnetic 
``minimal substitution'' 
$\nabla \to \nabla - \frac{iq}{\hbar c} A$. 
The basic idea, as formulated by Yang and 
Mills in 1954, leads
to the non-linear generalization of Electrodynamics that forms the
conceptual basis of the Standard Model, and can be motivated 
in the following simple terms. As we have seen, 
the electromagnetic gauge transformation
\begin{equation}
U=e^{\frac{iq\Lambda}{\hbar c}}
\end{equation}
is determined by a pure phase, that can be regarded as
a one-by-one unitary matrix, as needed, say, to describe the 
effect
of a rotation around the $z$ axis of three-dimensional Euclidean space on
the complex coordinate $x+iy$. Thus, one might well reconsider the whole 
issue of gauge invariance for an arbitrary rotation, or more generally for
$n\times n$ unitary matrices $U$. What would
happen then? First, the electrodynamic potentials would become matrices
themselves, while a gauge transformation would act on them as
\begin{equation}
\nabla - \frac{iq}{\hbar c} {\bf A}\ \to \ U \, \left(\nabla - \frac{iq}{\hbar c} {\bf A} \right) \, U^\dagger \ .
\end{equation}
Moreover, the analogs of the electric and magnetic fields would become
{\it non-linear} matrix-valued 
functions of the potentials, as can be seen repeating
the derivation in (\ref{commutator}) for a matrix potential ${\bf A}_\mu$,
for which 
\begin{equation}
\left[ \partial_\mu - \frac{iq}{\hbar c} {\bf A}_\mu, 
\partial_\nu - \frac{iq}{\hbar c} {\bf A}_\nu  \right] = - \frac{iq}{\hbar c} \,
\left(\partial_\mu {\bf A}_\nu - \partial_\nu {\bf A}_\mu -  
\frac{iq}{\hbar c}
[ A_\mu , A_\nu ] \right) = 
- \,   \frac{iq}{\hbar c} \ F_{\mu\nu} .
\label{commutator2}
\end{equation}
Notice that the matrix $({\bf A}_\mu)_i{}^j$ and the Christoffel
symbol $\left(\Gamma_\mu\right)_\nu{}^\rho$ are actually very similar
objects, barring from the fact the latter is not an independent field, but
a combination of derivatives of $g_{\mu\nu}$.

The resulting Yang-Mills equations
\begin{equation}
[\partial_\mu - \frac{iq}{\hbar c} {\bf A}_\mu,F^{\mu\nu}] = \frac{4 \pi}{c} \,
J^\nu \ ,
\end{equation}
to be compared with the more familiar Maxwell equations of Classical
Electrodynamics, 
contain indeed non linear (quadratic and cubic) terms that determine
the low-energy mutual interactions of gauge bosons.
For instance, the familiar Gauss law of Electrodynamics becomes
\begin{equation}
\nabla \cdot {\bf E} - \frac{iq}{\hbar c} \, ( {\bf A} \cdot {\bf E} -
{\bf E} \cdot {\bf A}) = 4 \pi \rho \ ,
\end{equation}
that can not be written in terms of ${\bf E}$ alone.
Notice also that the Yang-Mills analogs of ${\bf E}$
and ${\bf B}$ are not gauge invariant. Rather, under a gauge transformation
\begin{equation}
F_{\mu\nu} \to e^{ \frac{iq}{\hbar c} \Lambda} \ F_{\mu\nu} \ 
e^{ - \frac{iq}{\hbar c} \Lambda}
\end{equation}
so that the actual observables are more complicated in these non-Abelian
theories. An example is, for instance, ${\rm tr}(F_{\mu\nu} F^{\mu\nu})$,
while a more sophisticated, non-local one, is the Wilson loop
\begin{equation}
{\rm tr} \, P \, \exp\left(\frac{i q}{\hbar c} \oint_\gamma \,  A_\mu \, dx^\mu\right) \ ,
\end{equation}
where $P$ denotes path ordering, the prescription to order the powers of 
$A_\mu$ according to their origin along the path $\gamma$.
This non-Abelian generalization of the Aharonov-Bohm phase 
is of key importance in the problem of quark confinement.

The Standard Model indeed includes fermionic 
matter, in the form of quark and  lepton
fields, whose quanta describe three families of (anti)particles, but 
only the leptons are seen in isolation, so that the non-Abelian ${\rm SU(3)}$ 
color force is held responsible for the permanent confinement of quarks into
neutral composites, the hadrons. The basic interactions of quarks and leptons
with the gauge bosons are simple to characterize: as we anticipated,
they are determined by
minimal substitutions of the type $\nabla \to \nabla - \frac{iq}{\hbar c} 
{\bf A}$, but some of them violate parity or, in more technical language, are {\it chiral}. 
This fact 
introduces important constraints due to the possible occurrence of
{\it anomalies}, quantum violations of classical conservation laws. 
To give an idea of the
difficulties involved, it suffices to consider the Maxwell equations in
the presence of a current,
\begin{equation}
\partial_\mu \, F^{\mu\nu} = J^\nu \ . \label{curan}
\end{equation}
Consistency requires that the current be conserved, {\it i.e.} that
$\partial_\mu J^\mu=0$, but in the presence of parity violations 
quantum effects
can also 
violate current conservation, making (\ref{curan}) inconsistent.
Remarkably, the fermion content of the Standard Model passes this important
test, since all potential anomalies cancel among leptons and quarks.
\begin{figure}
\begin{center}
\psfig{figure=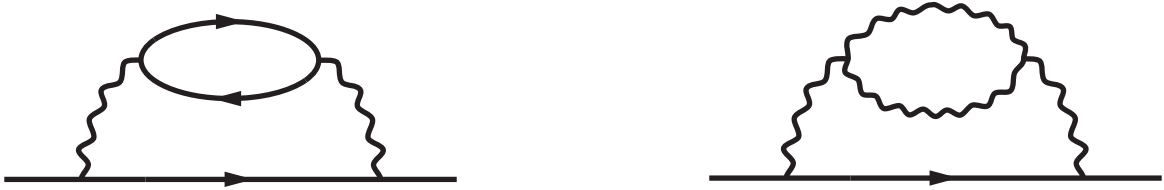}
\end{center}
\caption{The first diagram shows a typical contribution to the self-energy
of the electron. The virtual particle/anti-particle pair behaves as a 
small electric
dipole, thereby screening the electron charge. Turning to quarks and the 
strong interaction, in QCD
the diagrams of the first kind (where now the wavy line denotes a gluon,
rather than a photon) are accompanied by
additional ones of the second kind, since the gluons are themselves
charged. A direct calculation shows that the anti-screening effect wins, 
leading to {\it asymptotic freedom}. 
\label{fig:af}}
\end{figure}

Another basic feature of the Standard Model is related to the spontaneous
breaking of
the electro-weak symmetry, responsible for screening the weak force down
to very short distances, or equivalently for the masses of the 
$W^{\pm}$ and $Z$
bosons. This is achieved by the BEH mechanism, whereby the whole of space time
hosts a quartet of 
scalar fields responsible for the screening. Making a vector massive
costs a scalar field, that provides the longitudinal polarization of 
the corresponding
waves, so that three scalars are eaten up to build the $W^+$, $W^-$ and $Z$
bosons, while a fourth massive scalar
is left over: this is the Higgs, or more properly the BEH particle, whose
discovery would be a landmark event in Particle Physics.

After almost three decades, we are still unable to study 
the phenomenon of {\it quark confinement}
in fully satisfactory terms,
but we have a host of numerical evidence and simple
semi-quantitative arguments to justify our
expectations. Thus, in QED (see figure \ref{fig:af}) the uncertainty principle
fills the actual vacuum with virtual electron-positron pairs,
vacuum fluctuations that result in a
partial {\it screening} of a test charge. This, of course, can also radiate 
and absorb virtual
photons that, however, cannot affect the picture since they are uncharged.
On the other hand, the Yang-Mills vacuum (see figure \ref{fig:af}) 
is dramatically affected by the
radiation of virtual gauge bosons, that are charged and tend to
{\it anti-screen} a test charge. The end result of the two competing effects
depends on the relative weight of the two contributions, and the color
force in QCD is actually dominated by anti-screening. This has
an impressive consequence, known as {\it asymptotic freedom}: quark interactions
become feeble at high energies or short distances, as reflected in 
the experiments on deep inelastic scattering.
A naive reverse extrapolation would then appear to justify intense interactions
in the infrared, compatibly with the evident impossibility of finding
quarks outside hadronic compounds, but no simple quantitative proof of 
quark confinement 
has been attained to date along these lines. On the contrary, 
even if the weak interactions are also described by a Yang-Mills theory,
no subtle infrared physics is expected for them, compatibly with the 
fact that leptons are commonly seen in isolation: 
at scales beyond the Compton wavelength of the intermediate bosons,
$\ell_W \sim 10^{-16}\ cm$, the resulting forces are in fact 
screened by the BEH mechanism! 

\begin{figure}[h]
\begin{center}
\psfig{figure=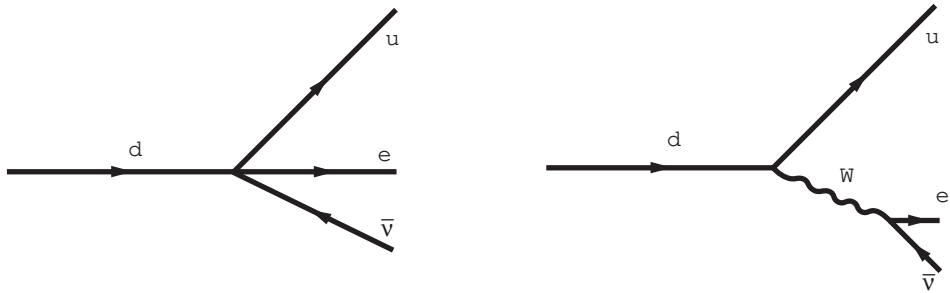}
\caption{At energies significantly below $100$ GeV, beta decay is well described by
a four-fermion interaction. On the other hand, at higher energies 
the interaction is resolved, 
or ``smeared out'', by the exchange of a $W$ boson.
\label{fig:fermi}}
\end{center}
\end{figure}

While more can be said about the Standard
Model, we shall content ourselves with these cursory remarks, with an 
additional comment on the nature of the spontaneous breaking. This ascribes
the apparent asymmetry between, say, the short-range weak interactions and the
long-range electromagnetic interactions to an asymmetry of the vacuum, much in the same way as the 
magnetization of a bar can be related to a proper hysteresis. As a result,
although hidden,
the symmetry is still present, and manifests 
itself in full power in high-energy
virtual processes, making the theory {\it renormalizable} like Quantum 
Electrodynamics is, a crucial result recognized
with the Nobel prize to G. 't Hooft and M. Veltman in 1999.
A by-product of the BEH mechanism is a simple relation between the Fermi 
constant and, say, the $W$ mass $M_W$
\begin{equation}
G_F \sim \frac{\hbar^3\alpha}{c\,M_W^2} \ ,
\end{equation}
with $\alpha$ a dimensionless number of the order of the QED 
fine-structure constant.
This reflects again the fact that the weak forces are completely screened 
beyond the Compton wavelength of their carriers, $\lambda_W \sim \frac{\hbar}
{c M_W}$, but an
equivalent, rather suggestive way of stating this result, is to note that
the growth of the effective fine-structure function
\begin{equation}
\alpha_F(E) \sim \frac{G_F \, E^2}{\hbar^3c^3} \ ,
\end{equation}
actually {\it stops} at the electro-weak scale 
$E_W \approx \sqrt{\hbar^3c^3}/\sqrt{G_F}$ to leave room to an essentially constant coupling.
This transition results from the emergence of new degrees of freedom that 
effectively smear out the local four-Fermi interaction into QED-like
exchange diagrams, as in figure \ref{fig:fermi}.

\begin{figure}[h]
\begin{center}
\psfig{figure=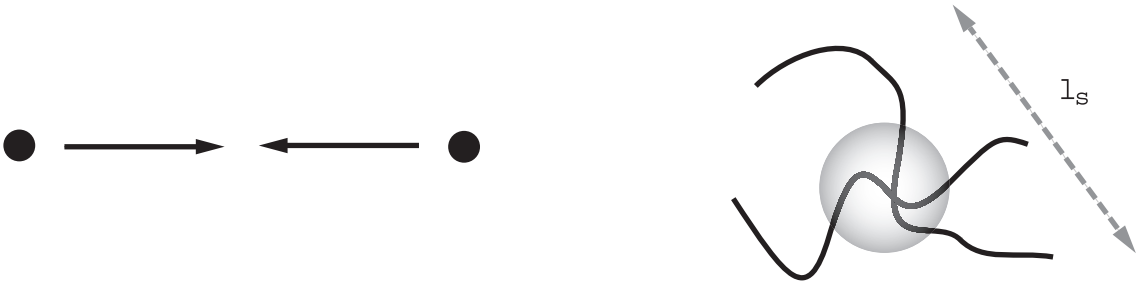}
\caption{A hard gravitational collision can be softened if point
particles are replaced by strings of length $\ell_s$. The energy is then
spread over their length, while at high energies
the effective gravitational coupling $\alpha_N(E)=G_N E^2/\hbar c^5$ is 
replaced by $\alpha_N(E)\times ((\hbar c /E)/\ell_s)^2$, according to the
fraction of the energy effective in the collision.
\label{fig:dilute}}
\end{center}
\end{figure}

\begin{figure}[h]
\begin{center}
\psfig{figure=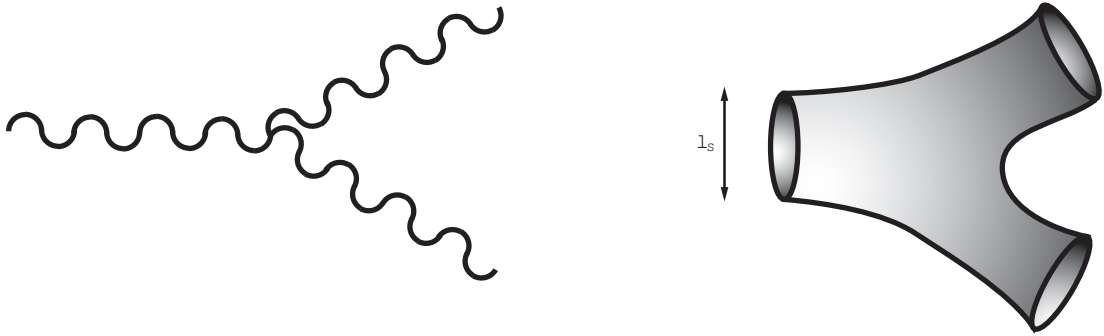}
\caption{Replacing point
particles with strings, a graviton three-point 
interaction (left) is ``smeared''
to such an extent that, in the resulting three-string vertex (right) even
no localized interaction point can be found anymore.
\label{fig:graviton}}
\end{center}
\end{figure}

Given these considerations, it is tempting and natural to try to
repeat the argument 
for gravity, constructing the corresponding dimensionless coupling,
\begin{equation}
\alpha_N(E) \sim \frac{G_N \, E^2}{\hbar \,c^5} \ .
\end{equation}
The relevant scale is now the Planck scale $E_{Pl} \approx 
10^{19} GeV$, but
the problem is substantially subtler, since now
energy itself is to be spread, and here is where strings come into play.
According to figures \ref{fig:dilute} and \ref{fig:graviton}, a  simple, if rather crude, argument to this effect 
is that if a pair of point masses experiencing a hard gravitational
collision are replaced with strings of length $\ell_s$, asymptotically 
only a fraction of their energies is effective in the interaction, so that
$\alpha_N(E)$ actually saturates to a finite limiting value, $G_N\hbar /\ell_s^2 c^3$.
This simple observation can be taken as the key motivation for strings in this
context, and indeed a detailed analysis shows that the ultraviolet problem of
gravity is absent in String Theory. A subtler issue
is to characterize what values of $\ell_s$
one should actually use, although naively the previous argument would 
lead to identify $\ell_s$ with the Planck length
$\ell_{Pl}\approx 10^{-33}cm$.

\section{From fields to strings}
 
This brings us naturally to strings, that clearly come in two varieties, 
{\it open} and {\it closed}. It is probably quite familiar that an ordinary 
vibrating string has an infinity of harmonics, that depend on the
boundary conditions at its ends, but whose frequencies are essentially
multiples 
of a fundamental tone. In a similar fashion, a single 
relativistic string has an
infinity of tones, naively related to an infinity of masses according to
\begin{equation}
m^2 \sim N \, \omega^2 \ , \qquad ( N \geq 1 )
\label{freqs}
\end{equation}
with $N$ an integer, and thus apparently
describes an infinity of massive particle species. 
There is a remarkable surprise, however: the dynamics
of strings requires a higher-dimensional Minkowski space and
typically turns the previous relation into
\begin{equation}
m^2 \sim (N - 1) \, \omega^2 \ ,
\end{equation}
so that string spectra actually
include massless modes, as needed to 
describe {\it long-range forces}. A more detailed analysis would reveal that 
open strings include massless
vectors, while closed strings include massless spin-2 fields. 
Therefore, not only is one softening the gravitational interactions by
spreading mass or energy, but one is also recovering without further ado
gauge bosons and gravitons from the string modes.

A closer look would reveal that strings can also describe
space-time fermions, with the chiral interactions needed in the Standard Model.
Their consistency, however, rests on a new mechanism, discovered by
M.B. Green and J.H. Schwarz, that supplements the ordinary anomaly 
cancellations at work in the Standard Model with the contributions
of new types of particles. In their simplest manifestation, these have
to do with a {\it two-form} field, a peculiar
generalization of the electrodynamic potential $A_\mu$ bearing an
antisymmetric pair of indices, so that $B_{\mu\nu}=-B_{\nu\mu}$. 
The corresponding field
strength, obtained as in Electrodynamics from its curl, is in this case
the {\it three-form} field 
$H_{\mu\nu\rho}= \partial_\mu B_{\nu\rho}+\partial_\nu B_{\rho\mu}+
\partial_\rho B_{\mu\nu}$. 
Two-form fields have a very
important
property: their basic electric sources are {\it strings}, just like the 
basic electric sources in the Maxwell theory are {\it particles}. 
Thus, in retrospect, a $B_{\mu\nu}$ field is a clearcut 
signature of an underlying string extension. A field of this type is
always present in the low-energy spectra of string models, but is absent in
the CJS supergravity that, for this reason, as stressed in the Introduction, 
bears no direct relation to strings.

We have already mentioned that there are apparently
several types of string models, all defined
in space times with a number of extra dimensions. At present,
the only direct way to describe their interactions is via a 
perturbative expansion.
Truly enough, this is essentially the case for the Standard Model as well,
but for strings we still lack somehow a way to go
systematically beyond perturbation theory. There is a framework, known
as String Field Theory, vigorously pursued over the years by a small
fraction of the community, and most notably by A. Sen and
B. Zwiebach, that is starting
to produce interesting information on the string vacuum state, but it is 
still a bit too early to give a fair assessment of its real potential in this
respect. Indeed, even the very 
concept of a string could well turn out to be provisional, a convenient
artifice to describe in one shot an infinity of higher-spin fields, much in 
the spirit of how a generating function in Mathematics allows one to 
describe conveniently in one shot 
an infinity of functions, and in fact String Theory
appears in some respect as a BEH-like phase of a theory with higher spins 
\footnote{The Standard Model contains particles of spin 1 (the gauge bosons),
1/2 (the quarks and leptons) and 0 (the BEH particle), and possibly of
spin 2 (the graviton), while the massive
string excitations have arbitrarily high spins.}. This is
another fascinating, difficult ad deeply related subject, pursued over the 
years mostly in Russia, and mainly by E. Fradkin and M. Vasiliev.

String Theory
allows two types of perturbative expansions. The first is regulated
by a dimensionless parameter, $g_s$, that takes the place in this 
context of the
fine-structure constants present in the Standard Model, while the second 
is a low-energy expansion, regulated by the ratio between typical energies
and a string scale $M_s \sim \hbar/c\,\ell_s$ related to the 
``string size'' $\ell_s$. In the
following we shall use interchangeably the two symbols $\ell_s$ and
$\alpha'=\ell_s^2$ to characterize the string size. A key
result of the seventies, due mainly to J. Scherk, J.H. Schwarz and T. Yoneya, 
is that in the low-energy limit the string interactions
embody both the usual gauge interactions of the Standard Model and
the gravitational interactions of General Relativity. Thus, to reiterate, 
String
Theory embodies by necessity long-range electrodynamic and gravitational
quanta, with low-energy interactions consistent with the Maxwell
(or Yang-Mills) and Einstein  equations.

The extra dimensions require that a space-time version of symmetry breaking be
at work to recover our four-dimensional world. The resulting framework draws
from the original work of Kaluza and Klein, and has developed into the
elegant and rich framework of Calabi-Yau compactifications, but some of
its key features can be illustrated by a simple example. To this end, let us
consider a massless scalar field $\phi$ that satisfies in five dimensions
the wave equation
\begin{equation}
\frac{1}{c^2}\, \frac{\partial^2 \phi}{\partial t^2} \ - \ \nabla^2 \phi \ -\ 
\frac{\partial^2 \phi}{\partial y^2} \ = \ 0 \ ,
\end{equation}
where the fifth coordinate has been denoted by $y$. Now suppose that $y$ 
lies on a circle of radius $R$, so that $y\sim y+2\pi R$, or, equivalently, 
impose periodic boundary conditions in the $y$-direction. One can then
expand $\phi$ in terms of a complete set of eigenfunctions of the circle
Laplace operator, plane waves with quantized momenta, writing
\begin{eqnarray}
\phi (x,y)=\sum_{n\in\IZ}\phi_n(x) \, e^{i\, n\,y/R} \ .
\end{eqnarray}
Plugging this expansion in the Klein-Gordon equation shows that,
from the four-dimensional viewpoint, the mode coefficients $\phi_n(x)$
describe independent fields with masses $n/R$, satisfying
\begin{equation}
\frac{1}{c^2}\, \frac{\partial^2 \phi_n}{\partial t^2} \ - \ \nabla^2 \phi_n 
\ + \ \frac{n^2}{R^2}\, \phi = 0 \ .
\end{equation}
At low energies, where the massive modes 
are frozen, the extra dimension is thus effectively screened and inaccessible,
since only quanta of the zero-mode field can be created.
Simple as it is, the example suffices to show that
the spectrum of massive modes reflects the features of the internal space,
in that it depends on the radius $R$. By a slight complication, for instance
playing with anti-periodic modes, one could easily see how even the numbers 
and types of low-lying modes present reflect in general the
features of the internal space. This is perhaps the greatest flaw in our 
current understanding: the four-dimensional manifestations of a given
string, and in particular the properties of its light particles, 
are manifold, since
they depend on the size and shape of the extra dimensions.
Let us stress that, while in the electro-weak breaking we dispose of a clear 
minimum principle that drives the choice of a vacuum, no general principle
of this type is available in the presence of gravity. Therefore, 
despite many efforts over the years, we have at present no 
clearcut way to make a dynamical choice between the available possibilities
and, as a result, we are still not in a position to give clearcut string 
predictions for low-energy parameters. Nonetheless,
these possibilities include, rather surprisingly, four-dimensional 
worlds with gauge and matter configurations along the lines of the 
Standard Model, although inheriting chiral interactions 
from higher dimensions would naively appear quite difficult. We may 
thus be driven to
keep an eye on a different and less attractive possibility, as with the 
old, ill-posed problem, of deriving from first principles
the sizes of the Keplerian orbits. 
As we now understand, these result from accidental initial conditions, and 
a similar situation for the four-dimensional string vacuum, while clearly 
rather disturbing, cannot be fairly dismissed at the present time.

Still, in moving to String Theory as the proper framework to
extend the Standard Model, it would be reassuring to foresee some sort 
of uniqueness in the resulting picture, at least in higher dimensions. 
Remarkably this was achieved, to 
a large extent, by the mid nineties, and we have now good reasons to
believe that all ten-dimensional superstring models are somehow 
equivalent to one another. 
The basic equivalences between the four superstring models of oriented
closed strings, IIA, IIB,
heterotic SO(32) and heterotic ${\rm E}_8 \times {\rm E}_8$, 
and the type I model of
unoriented closed and open strings, 
usually called 
{\it string dualities}, are summarized in figure \ref{fig:duality}.
The solid links, labeled by $T$ and $\Omega$, can be explicitly established
in string perturbation theory, while the additional dashed links rely on 
non-perturbative arguments that rest on the unique features of the 
low-energy ten-dimensional supergravity. We can now comment a bit on 
the labels, beginning with the $T$ duality.

\begin{figure}
\begin{center}
\psfig{figure=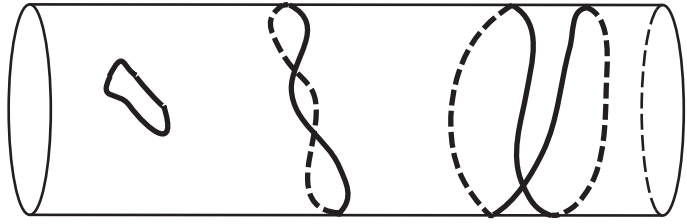}
\caption{Contrary to point particles, closed strings can wind in a non-trivial
way around a compact direction. here we show, from left to right, three 
examples of closed strings with winding numbers 0, 1 and 2, respectively.
\label{fig:compact}}
\end{center}
\end{figure}

When a particle lives in a circle, the de Broglie 
wave $e^{\frac{i p x}{\hbar}}$
can be properly periodic only if the momentum is quantized in units of the
inverse radius $R$, {\it i.e.} if $p=n \hbar/R$. We have already met the field
counterpart of this property above, when we have described how a massless
five-dimensional field would manifest itself to a four-dimensional observer 
as an infinite tower of massive fields. A closed string can 
also be endowed with a
center of mass momentum, quantized for the same reason in units of $1/R$,
and thus a single string spectrum would 
appear by necessity to a lower-dimensional 
observer as a tower of string spectra. However, a closed string can also wrap 
around the circle an arbitrary number of times, so that in fact a closed 
string coordinate admits expansions of the type
\begin{equation}
X(\sigma,\tau) = x \ + \ (2\alpha') \frac{m}{R} \, \tau \ + \ 2 n R \sigma \ + 
\ \frac{i}{2} 
\sqrt{2 \alpha'}\
\sum_{k \neq 0} \left(\frac{\alpha_k}{k}\, e^{-i2k(\tau-\sigma)} \ +\  
\frac{\tilde{\alpha}_k}{k} \, e^{-i2k(\tau+\sigma)}\right) \ ,
\label{closedcircle}
\end{equation}
where $\tau$ replaces in this context the ``proper time''
of particle dynamics while $\sigma$ labels the points of the string.
Notice that the third term implies that $X(\pi,\tau)=X(0,\tau) + 2 \pi n R$,
as pertains to a closed string winding $n$ times around a circle.
The spectrum of the string as seen from the uncompactified dimensions will
have the form,
\begin{eqnarray}
M^2= \frac{\hbar^2}{c^2}\left( \frac{m^2}{R^2}+ \frac{n^2R^2}{\alpha'{}^2}\right)+\cdots,
\label{spectrum}
\end{eqnarray}
where the dots stand for contributions due to the higher frequencies of the string (see e.g.
Eq. \ref{freqs}). While the first term in (\ref{spectrum})
is familiar from ordinary Quantum Mechanics, the second, that as we have 
seen reflects 
the possibility of having non-trivial windings, is new and intrinsically ``stringy''.
Notice that Eq. \ref{spectrum} displays a remarkable symmetry: one
cannot distinguish somehow between a string propagating on a circle of 
radius $R$
and another propagating on a circle with the ``dual'' radius $\alpha'/R\,$! 
We have actually simplified matters to some extent, 
since in general $T$-duality 
affects the fermion spectra of closed strings. Upon circle compactification,
it thus
maps the two heterotic models and the two type II models into one another,
providing two of the solid duality links in figure \ref{fig:duality}. 

The other solid link, labeled by $\Omega$, reflects an additional peculiarity,
the simultaneous presence of two sets of modes in a closed
string (the ``right-moving'' $\alpha$ and 
``left-moving'' $\tilde{\alpha}$ modes in Eq. \ref{closedcircle}). 
If a symmetry is present between
them, as is the case only for the type IIB model, one can use it to
combine states, but string consistency conditions require in general that new
sectors emerge. As a result, combining in this fashion
states of closed strings, one is generally led
to introduce open strings as well. This construction, now commonly called an
{\it orientifold}, was introduced long ago by one of the present authors
and was then widely pursued over the years at the University of Rome 
``Tor Vergata''. It links the type IIB and type I models in the diagram,
offering also new perspectives on the issue of string compactification.

The additional, dashed links in figure \ref{fig:duality},
are harder to describe in simple terms, but can be characterized as
analogues, in this context, of the electric-magnetic duality of Maxwell's
Electrodynamics.
It is indeed well-known that, in the absence of sources, 
the electric-magnetic duality transformations
${\bf E} \to {\bf B}$ and ${\bf B} \to - {\bf E}$ are a symmetry of
the Maxwell equations
\begin{eqnarray}
\nabla \cdot {\bf E} = 0 \qquad
\nabla \times {\bf E} = - \frac{1}{c}\, \frac{\partial {\bf B}}{\partial t}
\nonumber \\ 
\nabla \cdot {\bf B} = 0 \qquad\nabla \times {\bf B} =  \frac{1}{c}\, \frac{\partial {\bf E}}{\partial t} \ , \label{maxweqs}
\end{eqnarray}
but it is perhaps less appreciated that the symmetry can be extended to
the general case, at the expense of turning electric
charges and currents into their magnetic counterparts. Whereas these are
apparently not present in nature, Yang-Mills theories generically, but
not necessarily, predict
the existence of heavy magnetic poles, of masses $M \sim M_W/\alpha_e$, with
$\alpha_e$ a typical (electric) fine-structure constant. Thus, we might 
well have failed to see existing magnetic poles, due to their high
masses, about 100 times larger than those of the $W$ and $Z$ bosons! 
Actually, QED could be also formulated in terms of magnetic carriers, but
for Quantum Mechanics, that adds an important datum: the resulting magnetic 
fine-structure
constant $\alpha_m$ would be enormous, essentially the inverse of the 
usual electric one. More precisely, magnetic and electric couplings are 
not independent, but are related by Dirac quantization conditions,
so that
\begin{equation}
\alpha_e \sim \frac{1}{\alpha_m} \ ,
\end{equation}
and therefore it is the smallness of $\alpha_e$ that favors the usual
electric description, where the actual interacting electrons and photons 
are only mildly different from the corresponding free quanta, on which our
intuition of elementary particles rests.

In String Theory, the ``electric'' coupling is actually determined by
the vacuum expectation value of a ubiquitous massless scalar field, 
the {\it dilaton} (closely related to the Brans-Dicke scalar, a natural extension of
general relativity), according to
\be
g_s = e^{\langle \varphi \rangle} \ ,
\ee
and at this time we have no direct insight on $g_s$, that in general
could be space-time dependent. It is thus interesting to play with
these $S$ dualities, that indeed fill the missing gaps in 
figure \ref{fig:duality}. A surprise is that both
the type IIA and the ${\rm E}_8 \times {\rm E}_8$ heterotic models 
develop at strong coupling an 
additional dimension, invisible in perturbation 
theory, but macroscopic if $g_s$ is large enough. The emergence 
of the additional
dimension brings into the game the CJS supergravity, the unique
supergravity model in eleven dimensions, that however can 
not be directly related to strings: as we have stressed, 
it does not contain a $B_{\mu\nu}$ field, although it does contain 
a three-index field, $A_{\mu\nu\rho}$, related to corresponding
higher-dimensional solitonic objects, that we shall briefly return to
in the next section, the M2- and M5-branes.
This is the puzzling end of the story alluded to
in the Introduction: duality transformations of string models, that
supposedly describe the microscopic degrees of freedom of our world, 
link them to a supergravity model with no underlying string. This is 
indeed, in some respect, like ending up with pions with no clue on the 
underlying ``quarks''! This beautiful picture was contributed in the
last decade by many authors, including M. Duff, A. Font, 
P. Horava, C.M. Hull, L. Ibanez, D. Lust, F. Quevedo, A. Sen,
P.K. Townsend, and  most notably by E. Witten.

Let us conclude this section by stressing that a duality is a complete
equivalence between the spectra of two apparently distinct theories. We
have met one example of this phenomenon above, when we have discussed
the case of $T$ duality: winding modes find a proper counterpart in 
momentum modes, and vice versa. Now, in relating the heterotic
SO(32) model, say, to the type I string, their two sets of modes have no way
to match directly. For instance, a typical open-string coordinate
\begin{equation}
X = x + (2\alpha') \frac{m}{R} \, \tau + 2 n R \sigma + i \sqrt{2 \alpha'}\
\sum_{k \neq 0} \frac{\alpha_k}{k} e^{-ik\tau} \, \cos(k \sigma), 
\end{equation}
is vastly different from the closed-string expansion met above, since for
one matter it involves a single set of modes. How can a correspondence of
this type hold? We have already stumbled on the basic principle, when
we said that typically Yang-Mills theories also describe magnetic poles. These
magnetic poles are examples of {\it solitons}, stable localized blobs of energy
that provide apparently inequivalent descriptions of wave quanta,
to which we now turn.
\section{From strings to branes}

\begin{figure}
\begin{center}
\psfig{figure=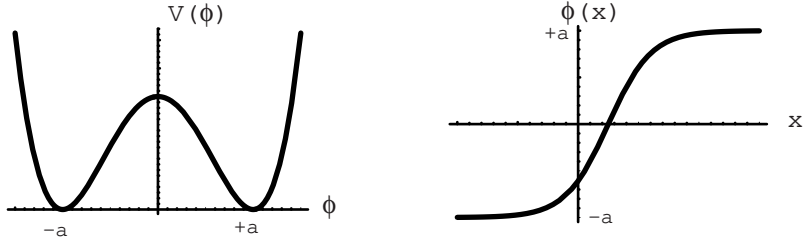}
\end{center}
\caption{A scalar field theory in $1+1$ dimensions (denoting the scalar
field as $\phi(t,x)$),
with a potential $V(\phi)=\lambda(\phi^2-a^2)^2$ (shown on the left)
admits a static, finite energy solution that interpolates 
between the two vacua $\phi=\pm a$ (shown on the right), known as a ``kink''. 
Its energy or mass is 
$m^3/12\lambda$, where $m$ is the mass of the elementary scalar field. 
In the perturbative regime, {\it i.e.} for small coupling $\lambda$, the 
kink is therefore very heavy.
\label{fig:kink}}
\end{figure}

A number of field theories admit solitonic solutions, 
blobs of energy whose shape is stabilized by non-linear
couplings.  A simple example is provided by the ``kink'',
that interpolates between the two minima of the potential 
shown in figure \ref{fig:kink}.
It can be regarded as a model for a wall separating a pair of
Curie-Weiss domains in a ferromagnet. Its stability can be argued by noticing
that any attempt to deform it, say, to the constant vacuum $\phi=a$ would
cost, in one dimension, an energy of the order of $L V(0)$, where $L$ is 
the size of the region where the field theory lives 
and $V(0)$ is the height of the potential barrier. 
For a macroscopic size $L$
this becomes an infinite separation, and the solution is thus 
stable. Moreover, its
energy density, essentially concentrated in the transition region,
results in a {\it finite} total energy, $E=m^3/12\lambda$, where $m$
denotes the mass of the elementary scalar field, defined expanding around
one of the minima of the potential, $\phi=\pm a$. This energy defines the 
mass of the soliton and, as anticipated, blows up in the limit of 
small coupling $\lambda$.
The `t Hooft-Polyakov monopole works, in three dimensions, along similar 
lines: any attempt to
destroy it would cost an infinite energy: as is usually said, these objects are
{\it topologically stable}, and in fact their stability can be ascribed to the
conservation of a suitable (topological) charge, that for the
monopole is simply its magnetic charge.
A further feature of solitons is that their energy is proportional to
an inverse power of a coupling constant, as we have seen for the kink. 
This is simple to understand in general terms: 
the non-linear nature of the field equations is
essential for the stability of solitons, that therefore should disappear in
the limit of small coupling! 

A localized distribution of energy and/or charge is indeed 
a modern counterpart of our classical idea of a particle. It is probably
familiar that an electron has long been
modeled in Classical Electrodynamics, in an admittedly {\it ad hoc}
fashion, as a spherical shell
with a total charge $e$ and a finite radius $a$, associating  
the resulting electrostatic energy
\begin{equation}
E \sim \frac{e^2}{a}
\end{equation}
with the electron mass. In a similar fashion, the
localized energy distribution of a soliton is naturally identified 
with a particle, just like an energy distribution localized along 
a line is naturally identified with an infinite string, while its higher
dimensional analogs define generalized {\it branes}. Thus, for instance,
the ``kink'' describes a particle in 1+1 dimensions, a string in 1+2
dimensions, where the energy distribution is independent of a spatial
coordinate, and a domain wall or {\it two-brane} in 1+3 dimensions, 
where the energy 
distribution is independent of two spatial coordinates.
These are therefore new types of ``quanta'', somehow missed by our
prescription of reading particle spectra from free wave equations. 
Amusingly
enough, one can argue that the two descriptions of particles are only
superficially different, while the whole picture is well fit with Quantum 
Mechanics. The basic observation is that these energy blobs have typically a 
spatial extension 
\begin{equation}
\Delta \sim \frac{\hbar}{M c} \ , \label{size}
\end{equation}
where $M$ denotes a typical mass
scale associated to a BEH-like phenomenon, since they basically arise from
regions where a transition between vacua takes place, typically of
the order of the Compton wavelength (\ref{size}). In addition, the energy
stored in these regions, that determines the mass of the soliton, is
\begin{equation}
M_{sol} \sim \frac{M}{\alpha} \ , \label{msol}
\end{equation}
with $\alpha$ a typical fine-structure constant. At weak coupling 
(small $\alpha$) we
have quantitative means to explore further the phenomenon, but 
$M_{sol} \gg M$, so that the
Compton wavelength of the soliton is well within its size. In other words, in
the perturbative region the soliton is a {\it classical} object. 
On the other hand, in the strong-coupling limit (large $\alpha$)
the soliton becomes light while its
Compton wavelength spreads well beyond its size, so that its inner structure
becomes immaterial: we are then back to something very similar in all respects
to an ordinary quantum.
\begin{figure}
\begin{center}
\psfig{figure=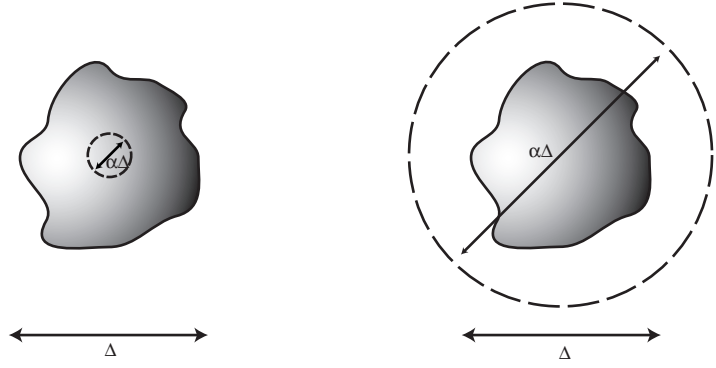}
\caption{An ``artistic'' impression of a soliton of size $\Delta$. 
The dashed circle depicts its Compton wavelength $\lambda_C\sim \alpha\Delta$,
with $\alpha$ a dimensionless fine-structure constant. The first figure
refers to the weak-coupling regime, where $\alpha$ is very small and, as a
result, the Compton wavelength $\lambda_C$ is much smaller than the soliton 
size $\Delta$. In this regime, the
soliton can be regarded as a classical object.
The second figure refers to the strong-coupling regime, where $\alpha$ is 
very large and, as a result, the
Compton wavelength $\lambda_C$ is much larger than the soliton size $\Delta$. 
In this regime, the soliton can be regarded as an ordinary light 
quantum without any inner substructure.
\label{fig:soliton}}
\end{center}
\end{figure}
Solitons are generally interacting objects. For instance, magnetic poles
typically experience the magnetic dual of the usual Coulomb force. This is
reflected in the fact that, being solutions of non-linear equations, they
can not be superposed. In special cases, their mutual forces can happen
to cancel, and then, quite surprisingly, the corresponding 
non-linear field equations allow a superposition of different solutions.
This is a typical state of affairs in supersymmetric theories, realized when
special inequalities, called ``BPS bounds'', are saturated.

The $T$ and $S$ dualities discussed in the previous section
can also be seen as maps between ordinary quanta and solitons.
The former, simpler case, involves the interchange of 
momentum excitations with winding modes that,
as we have stressed, describe topologically inequivalent closed-string 
configurations on a circle, while the latter rests on similar 
operations involving solitons in space time. These, as in the two examples
we have sketched in this section, can be spotted from the field
equations for the low-energy string modes, but some of their features
can be discussed in simple, general terms.
To this end, let us begin by rewriting the (1+3)-dimensional 
Maxwell equations (\ref{maxweqs}) in covariant notation, while extending them
to the form
\begin{eqnarray}
\partial_\mu \, F^{\mu\nu} &=& \frac{4 \pi}{c} \, J_e^\nu \ , \\ 
 \varepsilon^{\mu\nu\rho\sigma} \, \partial_\nu \, F_{\rho\sigma} &=& 
\frac{4 \pi}{c} \, {J}_m^\nu \ , \label{fnl}
\end{eqnarray}
where, in addition to the more familiar electric sources $J_e$, 
we have also introduced magnetic sources $J_m$, that affect 
the Faraday-Neumann-Lenz induction law 
and the magnetic Gauss law. Notice how a current $J_e^\mu$ is naturally 
borne by particles,
with $J_e^\mu \sim q \, u^\mu$ in terms of their charge and 
four-velocity. In $D > 4$, however, the $\varepsilon$
tensor carries $D$ indices, and consequently  $J_m$ carries in
general $D-3$ indices, while its sources are extended objects defined via
$D-4$ Lorentz indices. Thus, a magnetic pole is a particle in four
dimensions as a result of a mere accident. In six dimensions, for instance,
the magnetic equations would become
\begin{equation}
 \varepsilon^{\mu\nu\rho\sigma\tau\lambda} \, \partial_\sigma \, 
F_{\tau\lambda} =  \frac{4 \pi}{c} \, \tilde{J}_m^{\mu\nu\rho} \ ,
\end{equation}
so that by the previous reasoning a magnetic
pole would bear a pair of indices, as pertains to a surface. 
In other words, it would be a two-brane. The argument can be
repeated for a general class of tensor gauge fields, 
$B_{\mu_1 \cdots \mu_{p+1}}$, in $D$ dimensions: their electric sources 
are $p$-branes, while the corresponding magnetic sources are $(D-4-p)$-branes.
These tensor gauge 
fields are typically part of low-energy string spectra, while the
corresponding ``electric'' and ``magnetic'' poles show up as solutions of the 
complete low-energy equations for the string modes. As we have seen, they
define new types of ``quanta'' that are to be
taken into account: in fact, ``branes'' of this type are the missing states
alluded to at the end of the previous section!

As stressed by J. Polchinski, a peculiarity of String Theory makes 
some of the ``branes'' lighter
than others in the small-coupling limit and, at the same, 
simpler to study. The first
feature is due to a string modification of Eq. \ref{msol}, that for these
``D-branes'' happens to depend on $\sqrt{\alpha} \sim g_s$, 
rather than on $\alpha$,
as is usually the case for ordinary solitons. The second feature is related 
to the possibility of
{\it defining} String Theory in the presence of D-branes
via a simple change of boundary conditions at the string ends. 
In other words, D-branes absorb and radiate strings. In analogy with
ordinary particles, D-branes can be characterized by a {\it tension} (mass per
unit volume) and a {\it charge}, that defines their coupling to
suitable tensor gauge fields. While their dynamics is
prohibitively complicated, in the small coupling limit they
are just rigid walls, and so one is effectively studying some sort of 
Casimir effect induced by their presence. The idea is hardly new: for instance,
the familiar Lamb shift of QED is
essentially a Casimir effect induced by the atom. What is
new and surprising in this case, however, is that the 
perturbation theory around
D-branes can be studied in one shot for the whole string spectrum. In other
words, for an important class of phenomena that can be associated to 
D-branes, a {\it macroscopic} analysis of the corresponding field 
configurations
can be surprisingly accompanied by a {\it microscopic} analysis of their 
string fluctuations. This is what makes D-branes 
far simpler than other string
solitons, for instance the M5-brane, on which we have very 
little control at this time. The mixing of left and right closed-string
modes met in the discussion of orientifolds in the context of
string dualities can also be given a
space-time interpretation along these lines: it is effected by
apparently non-dynamical ``ends of the world'',
usually called O-planes. There is also an interesting possibility, well
realized in perturbative open-string constructions: while branes, being
physical objects, are bound to have a positive tension, one can 
allow different types of O-planes, with both negative and
positive tension. While the former are typical ingredients of
supersymmetric vacua, the latter can induce interesting mechanisms of 
supersymmetry breaking that we shall mention briefly in the
next section.

\section{Some applications}

The presence of branes in String Theory provides new perspectives on a
number of issues of crucial conceptual and practical import. In this
section we comment briefly on some of them, beginning with the
amusing possibility that our universe be associated to
a collection of branes, and then moving on to brief discussions of black hole
entropy and color-flux strings.

\subsection{Particle Physics on branes?}

\begin{figure}
\begin{center}
\psfig{figure=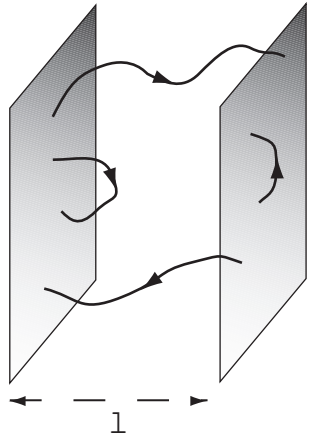}
\caption{A system consisting of two parallel D-branes, to which 
oriented open strings 
can attach in four different ways. The masses of the gauge fields 
associated to the four types of open 
strings are proportional to the shortest distances between 
the branes they connect. When the mutual distance $l$ between the branes 
is not zero,
one is thus describing two massless gauge fields, with a corresponding
unbroken $U(1)\times U(1)$
symmetry, and two additional massive $W$-like fields.
On the other hand, when $l\rightarrow 0$ the 
two $W$-like fields become massless as well,
while the gauge symmetry enhances to $U(2)$. In this geometric 
setting for the BEH mechanism, the Higgs scalar
describes the fluctuations of the branes relative to one another, 
while its vacuum value defines their relative distance $l$.
\label{fig:2branes}}
\end{center}
\end{figure}

One is now confronted with a fully novel situation: as these
``branes'' are extended objects, one is naturally led to investigate the 
physics of their interior or, in more pictorial terms, the physics as 
seen by an observer living on them. To this end, it is necessary to study
their small oscillations, that define the light fields or, from what we
said in the previous sections, the light species of particles seen by the
observer. These will definitely include the scalars that describe small
displacements of the ``branes'' from their equilibrium positions, and possibly
additional light fermionic modes. A surprising feature of D-branes is that 
their low-energy spectra also include {\it gauge fields}. Both scalars and
gauge fields arise from the fact that open strings end on D-branes 
(seen from the brane, their intersections are point-like), and are in 
fact associated to string fluctuations transversal or longitudinal 
to the branes, respectively. 
In addition, when several branes coincide
non-Abelian gauge symmetries arise, as summarized in figure \ref{fig:2branes}.
In equivalent terms, the mutual displacement of branes provides a geometric
perspective on the BEH mechanism. Moreover, the low-energy dynamics 
of gauge fields on a $Dp$-brane is precisely of the Yang-Mills type, but at 
higher energies interesting stringy corrections come into play. While a proper
characterization of the general case is still an open problem,
in the Abelian case of a single D-brane, and in the limit of slowly 
varying electric and magnetic fields, String Theory recovers 
a beautiful 
action proposed  in the 1930's by Born and Infeld to solve the singularity 
problem of a classical point-like electric charge, as originally shown by 
E. Fradkin and A. Tseytlin. Let us explain briefly this point.
Whereas in the usual Maxwell formulation the resulting Coulomb field
\begin{equation}
{\bf E} = \frac{q}{r^2} \ {\bf \hat{r}} \label{maxwellcharge} \ ,
\end{equation}
where ${\bf \hat{r}}$ denotes the unit radial vector,
leads to an infinite energy, in String Theory the Maxwell 
action for the static case is modified, and takes the form
\begin{equation}
- \frac{1}{2} \, {\bf E}^2 \to \frac{1}{2\pi\alpha'} \ 
\sqrt{1 - 2\pi\alpha'\,  {\bf E}^2}
\end{equation}
so that Eq. (\ref{maxwellcharge}) is turned into
\begin{equation}
{\bf E} = \frac{q}{\sqrt{r^4+ (2\pi\alpha')^2}} \, {\bf \hat{r}} \ .
\end{equation}
As a result, the electric field strength saturates to $\frac{q}{2\pi\alpha'}$, 
much in the same way as the speed of 
a relativistic particle in a uniform field saturates to the speed of light $c$,
an analogy first stressed in this context by C. Bachas.
Thus, once more String Theory appears to regulate divergences, as we have
already seen in connection with the ultraviolet problem of gravity.

Summarizing, the world volume of a collection of D$p$-branes is by
construction a $(p+1)$-dimensional space that contains in principle 
the right types of light fields to describe
the particles of the Standard Model. This observation has changed our whole
perspective on the Kaluza-Klein scenario, is at the heart of
current attempts to model our universe as a collection of 
intersecting D-branes, and 
brings about a novelty that we would like to comment briefly upon. The issue
at stake is, again, the apparently unnatural hierarchy between the electro-weak
and Planck scales, on which this scenario offers a new geometric perspective,
since in a ``brane world'' gauge and matter interactions are confined to the
branes, while gravity spreads in the whole ambient space. One can thus 
provide a different explanation for the weakness of gravity: most of 
its Faraday lines
spread in the internal space, and are thus simply ``lost'' for a brane 
observer. This is the essence of a proposal made by I. Antoniadis, N.
Arkani-Ahmed, S. Dimopoulos and G. Dvali, that has stimulated a lot
of activity in the community over the last few years.
For instance, with $n$ extra circles of radius $R$ one would find that a 
$(4+n)$-dimensional Newton constant $G_{4+n}$ for bulk gravity induces
for two point-like masses on the brane an effective Newton constant
$1/G_4 \sim R^n/G_{4+n}$. This result can be obtained adding the 
contributions of the extra circles or, more simply, purely on dimensional
grounds. Playing with the size $R$, one can start with
$G_{4+n} \sim \left(1/TeV\right)^{2+n}$ and end up with the conventional 
$G_4 \sim 1/(10^{19} GeV)^2$, if $R \sim 10^{32/n} \times 10^{-4} fm$,
so that if $n \geq 2$ the resulting 
scenario is not obviously excluded. The phenomenon would
manifest itself as a striking change in the power law for the Newton force 
(\ref{newtforc}),
that for $r < R$ would behave like $1/r^{2+n}$, a dramatic effect indeed,
currently investigated by a number of experimental groups at scales 
somewhat below the millimeter. In a similar fashion, one can also conceive
scenarios where the string size $\ell_s$ is also far beyond the Planck
length, but a closer inspection shows that in all cases
the original hierarchy problem has been somehow rephrased in geometrical, 
although possibly
milder, terms: all directions parallel to the world brane should be far below
the millimeter, at least ${\cal O}(10^{-16} \ cm)$, if no new phenomena are to
be present in the well-explored gauge interactions of the Standard Model at
accessible energies, so that a new hierarchy emerges between longitudinal
and transverse directions. The literature also contains interesting
extensions of this scenario with infinitely extended curved internal
dimensions, where gravity can nonetheless be localized on branes, 
but this simpler case should suffice to give a flavor of the potential role
of branes in this context.

It is also possible to complicate slightly this picture to allow for the
breaking of supersymmetry. To date, we have only one way to introduce
supersymmetry breaking in closed strings working at the level of the full 
String Theory, as opposed to
its low-energy modes: Bose and Fermi fields can be given different harmonic
expansions in extra dimensions. For instance, referring to the case of 
Section 3, if along an additional circle Bose fields are periodic while Fermi
fields are anti periodic, the former inherit the masses $k/R$, while the
letter are lifted to $(k+1/2)/R$, with supersymmetry broken at a scale
$\Delta M \sim 1/R$. This is the Scherk-Schwarz mechanism, first fully
realized in models of oriented closed strings by S. Ferrara, K. Kounnas,
M. Porrati and F. Zwirner, following a previous analysis of R. Rohm.
Branes and their open strings, however, allow new possibilities, 
known in the literature, respectively, as ``brane supersymmetry'' and 
``brane supersymmetry breaking'', that we would like to briefly comment
upon. Of course, the mere presence of branes, 
extended objects of various dimensions, breaks some space-time symmetries,
and in fact one can show that a single brane breaks at least half of the
supersymmetries of the vacuum, but more can be done by suitable combinations
of them. Thus, the first mechanism
follows from the freedom to use, in the previous construction, 
directions parallel or transverse to the ``brane world'' to separate Fermi and
Bose momenta. While momenta along parallel
directions reproduce the previous setting, orthogonal ones in principle
{\it can not} separate brane modes. However, a closer inspection
reveals that this is only true for the low-lying excitations, while the
massive ones, affected by the breaking, feed it via radiative
corrections to the low-lying modes, giving rise a gravitational analogue
of the ``see-saw'' mechanism, with $\Delta M \sim \sqrt{G_N}/ R^2$. 
\begin{figure}
\begin{center}
\psfig{figure=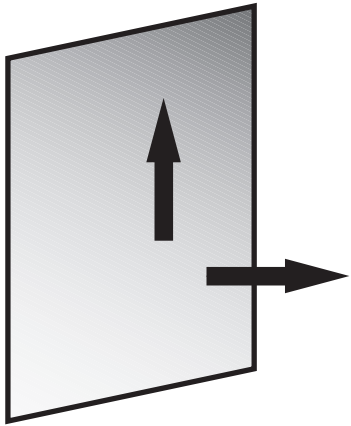}
\caption{Brane supersymmetry and the gravitational see-saw. If the momenta
of Bose and Fermi brane fields are separated along a direction orthogonal
to the D-brane, only the massive brane excitations feel the effect, that is 
then transmitted via radiative corrections to the low-lying excitations,
relevant for the Standard Model physics.
\label{fig:seesaw}}
\end{center}
\end{figure}
Finally, the second mechanism can induce supersymmetry breaking in our world 
radiatively from other non-supersymmetric branes, with the interesting 
possibility of attaining a low vacuum energy in the observable world.

By and large, however, one is again led to a puzzling end: a sort
of ``brane chemistry'' allows one to concoct an observable world out of these
ingredients, much in the spirit that associates chemical compounds to 
the basic elements of the Periodic Table, and eventually to electrons and
nuclei. However, the problem alluded to in the previous Sections is still 
with us: we have presently no plausible way of selecting a 
preferred configuration to connect String Theory to our low-energy
world, although one can well construct striking realizations of the 
Standard Model on intersecting branes, as first shown by the string groups
at the Universidad Autonoma de Madrid and at the Humboldt University in Berlin.
\begin{figure}
\begin{center}
\psfig{figure=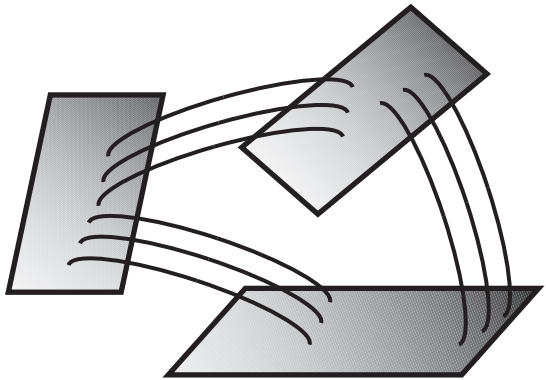}
\caption{The world as a collection of interacting branes. Amusingly, we 
have come all this way from our matter seen 
as a collection of point particles, recovering
at large scales something quite reminiscent of 
our starting point.
\label{fig:braneworlds}}
\end{center}
\end{figure}

\subsection{Can strings explain Black Hole Thermodynamics?}

As we have stressed, D-branes can be given a {\it macroscopic} description as
solutions of the non-linear field equations for the light string modes, and
at the same time a {\it microscopic} description as emitters and absorbers
of open strings. If for a black hole both descriptions were
available, one would be naturally led to regard the open string degrees of 
freedom as its own excitations. This appears to provide a new perspective on 
a famous
result of S. Hawking, that associates to the formation of a black hole a 
blackbody spectrum of radiation at a characteristic
temperature $T_H$. Since, as originally stressed by J. Bekenstein, the 
resulting conditions for the mass variation of
the hole have the flavor of Thermodynamics, D-branes offer
the possibility of associating to this Thermodynamics a corresponding
Statistical Mechanics, the relevant microstates being their own excitations.

In the presence of a static isotropic source of mass $M$ at the origin, 
the  Minkowski line element is deformed to
\begin{eqnarray}
ds^2=\left(1-\frac{2G_N M}{r\,c^2}\right)c^2dt^2-\left(1-\frac{2G_N 
M}{r\,c^2}\right)^{-1}dr^2-
r^2d\Omega^2.
\label{schwa}
\end{eqnarray} 
This expression holds {\em outside} 
the source, while the special value of the radial coordinate
$r_h=2G_NMc^{-2}$ corresponds to the {\it event horizon}, that can
be characterized as the minimum sphere centered at the origin that is
accessible to
a far-away observer. For most objects $r_h$ lies deep inside 
the source itself ({\it e.g.} for the sun $r_h \approx 3\, Km$, 
to be compared with the solar radius $R\approx 10^6 \, Km$), where 
Eq.~\ref{schwa} is not valid anymore, but one can conceive a 
source whose radius is inferior 
to $r_h$, and this is called a {\it black hole}: according to classical 
General 
Relativity, any object coming from outside and 
crossing the horizon is trapped inside it forever. 
Over the past decade, astrophysical observations have given strong, if
indirect, clues that black holes are ubiquitous in our universe.

As anticipated, however, Hawking found that black holes are not really black
if Quantum Mechanics is properly taken into account. Rather, quantizing a 
Field Theory in a background containing a black hole, he showed that to
an external observer the hole appears to
radiate as a black body with temperature
\begin{eqnarray}
T_H\ =\ \frac{c^3\, \hbar}{8\pi k_B G_N M}\ , \label{temp}
\end{eqnarray}
where $k_B$ denotes Boltzmann's constant.
This amazing phenomenon that can be made plausible by noting
that a virtual particle-antiparticle pair popping up in the neighborhood of 
the horizon can have such a dynamics that one of the two crosses the 
horizon, while the other, forced by energy conservation to materialize 
as a real particle, will do so absorbing and carrying away
part of the gravitational energy of the black hole.
In analogy with the second law of Thermodynamics, given the temperature $T_H$
one can associate to a black hole an {\it entropy}
\begin{eqnarray}
\frac{1}{k_B}\; S_H\ =\ \frac{4\pi G_N M^2}{c\hbar}\ = \
\frac 1 4 A_H \, \ell_{Pl}^{-2} \ , \label{schwarz}
\end{eqnarray}
where $A_H$ is the area of the horizon and
$\ell_{Pl}$ is the Planck length $\ell_{Pl}=\sqrt{G_N \hbar/c^3}\approx  
\ 10^{-33}\,cm$, that we have repeatedly met in the previous sections. 
This expression, known as the Bekenstein-Hawking formula,
reflects a universal behavior: the 
entropy of any black hole is one quarter of the area of its horizon in 
Planck units. 

Several questions arise, that have long puzzled many experts:
\begin{itemize}
\item As anything crossing the horizon disappears leaving only 
thermal radiation behind, the S-matrix of a system containing a black 
hole seems not unitary anymore, thus violating a basic tenet of Quantum
Mechanics. This is known as the {\it information paradox}.
\item Entropy is normally a measure of the degeneracy of microstates $\Sigma$
in some underlying 
microscopic description of a physical system, determined by Boltzmann's formula,
\begin{equation}
S = k_B \, \log \Sigma \ .
\end{equation}
Since the entropy (\ref{schwarz})
of a black hole is naturally a huge number,
how can one exhibit such a wealth of microstates?
\item Eq.~\ref{temp} clearly shows that the more mass is radiated away 
from the black hole, the hotter this becomes. What is then the endpoint of 
black hole evaporation?
\end{itemize} 

Within String Theory there is a class of black holes where 
these problems
can be conveniently addressed, the so-called extremal black holes, that 
correspond to BPS objects in this context. The 
simplest available example is provided by 
a source that also carries an electric charge $Q$. The coupled 
Maxwell-Einstein 
equations would give in this case the standard Coulomb potential for 
the electric field, together with the modified line element $(G_N M^2 > Q^2)$
\begin{eqnarray}
ds^2=\left(1-\frac{2G_NM}{r\,c^2}+\frac{G_N Q^2}{r^2\,c^4}\right)dt^2-\left(1-\frac{2G_NM}{r\,c^2}+
\frac{G_N Q^2}{r^2\,c^4}\right)^{-1}dr^2- r^2d\Omega^2,
\label{schwa2}
\end{eqnarray}
that generalizes Eq.~\ref{schwa}. Notice that the additional terms 
in (\ref{schwa2}) have
a nice intuitive meaning: $Q^2/2r$ is the electrostatic 
energy introduced by the charge in the region beyond $r$, and this
contribution gives rise to a repulsive gravitational effect. 
The event horizon, defined again as the smallest sphere surrounding the hole
that is accessible to a far-away observer, would now be
\begin{eqnarray}
r_H=c^{-2}\left(G_NM+\sqrt{(G_NM)^2-G_N Q^2}\right).
\end{eqnarray}
A source with a radius smaller than $r_H$ would be a Reissner-Nordstrom 
black hole,
with temperature and entropy given by
\begin{eqnarray}
T_H&=&\frac{c^3\hbar\sqrt{(G_NM)^2-G_N Q^2}}{2\pi k_B (G_NM+\sqrt{(G_NM)^2-G_N Q^2}\,)^2},\nonumber\\
\frac{S_H}{k_B}&=& \frac{\pi}{c\hbar G_N}(G_NM+\sqrt{(G_NM)^2-G_N Q^2}\,)^2  =\frac 1 4 A_H \, l_p^{-2}. \label{reissner}
\end{eqnarray}
For a given value of $Q$, if $M\rightarrow Q/\sqrt{G_N}$ 
the temperature vanishes, so that the black hole behaves somehow in this 
limiting (BPS) case as if it 
were an elementary particle. Such a black hole is called extremal: its mass 
is tuned so that the tendency to 
gravitational collapse is precisely balanced by the electrostatic 
repulsion. This limiting case entails a manifestation of the phenomenon alluded
to in Section 4: although the Maxwell-Einstein equations are highly non linear,
one can actually superpose these extremal solutions.

Extremal black holes of this type can be described in 
String Theory in relatively simple terms. One of the simplest configurations  
involves the type IIB string theory compactified on a 
5-dimensional torus, together with a D5-brane and a D1-brane wrapped $n_5$ 
and $n_1$ times respectively around the torus. This BPS 
configuration is characterized by two topological numbers, $n_1$ and $n_5$,
but one needs a slight complication of it since, being the only BPS 
state with these charges, it leads to a vanishing entropy, 
consistently with Eq. \ref{reissner}. 
However, suitable excitations, involving 
open strings ending on the D-branes and wrapping in various ways around 
the torus, are also BPS and can be characterized by a single
additional quantum number, $n_e$. 
Many open string configurations now correspond to a given value of $n_e$, 
and counting them one can obtain a {\it microscopic} estimate of the entropy. 
One can then turn to IIB supergravity on the 5-torus, constructing
a BPS solution of its field equations that involves the {\it three}
charges mentioned above, to calculate its event 
horizon, its temperature and finally to obtain the corresponding 
{\it macroscopic} estimate of the entropy. The exact agreement between the
two estimates is then striking. Since this original example 
was discussed by A. Strominger and C. Vafa, 
many other black hole configurations were studied, while the analysis was 
successfully extended to nearly extremal ones. These results, however, rely
heavily on supersymmetry, and serious difficulties are met in attempts to
extend them to non-supersymmetric black holes.

The analysis of nearly extremal black holes 
also appears to provide a clue on the 
information paradox. Studying a configuration slightly away from 
extremality, it was indeed 
found that Hawking radiation can be associated to the {\it annihilation} of 
pairs of open strings, 
each ending on a D-brane, that give rise to open strings remaining 
on the brane and to closed strings leaving it. The resulting radiation turns 
out to be exactly thermal, while temperature and radiation rate 
are in perfect agreement with a Hawking-like calculation. 
Almost by construction, this process is unitary, and so the
information that seemed lost appears to be left in the D-branes.
\begin{figure}
\begin{center}
\psfig{figure=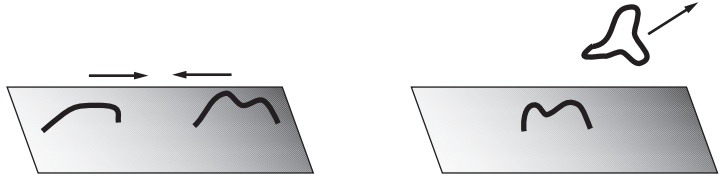}
\caption{The D-brane picture of Hawking radiation. A pair of open
strings collide, giving rise to a closed string that leaves the brane.
As a result, Hawking radiation reaches the bulk via the emission of
closed strings. 
\label{fig:bh}}
\end{center}
\end{figure}

\subsection{AdS/CFT: strings for QCD mesons, or is the universe a hologram?}

In the previous section we saw that the entropy of a black hole is proportional
to the area of its horizon. This is remarkable, since one can argue that 
black holes maximize the entropy. Indeed, assume for a moment that one 
managed to construct a physical system in a given volume $V$ with 
a mass $M-\delta M$ slightly inferior to that of a black hole whose horizon 
spans the surface surrounding $V$, but with an entropy 
$S+\delta S$ slightly larger than that of the
black hole. Throwing in a bit matter would then create a black hole 
while simultaneously 
lowering the entropy, thereby violating the fundamental law of Thermodynamics.
This observation led 't Hooft to propose the {\it 
Holographic Principle}: in a complete
theory of quantum gravity, it should be possible to describe the
physics of a certain region of spacetime in terms of degrees of freedom 
living on the surface surrounding it, while the information stored should be 
limited to roughly one bit per Planck area unit.
\begin{figure}
\begin{center}
\psfig{figure=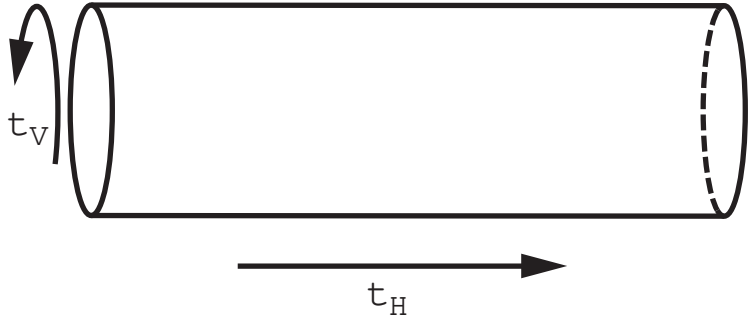}
\caption{A surprising equivalence in String Theory. An annulus diagram can be
regarded as a loop diagram for open strings (with vertical time $t_V$), or 
equivalently as a closed string tree diagram (with horizontal time $t_H$). 
\label{fig:openclosed}}
\end{center}
\end{figure}

Over the past few years, concrete realizations of the Holographic principle 
have been constructed, most dramatically in the context of the
so-called AdS/CFT correspondence. In its simplest form, this arises if
the type IIB string theory is defined in a ten-dimensional 
space-time with the topology of a five-dimensional sphere ($S^5$) times a
five-dimensional anti-de Sitter space ($AdS_5$), a non-compact manifold 
whose boundary can be identified
with four-dimensional Minkowski space. 
This geometry describes the region around
the horizon for a stack of $n$
D3-branes, that in the large-$n$ limit actually
invades the whole of space time.
On the one hand there are therefore 
D-branes, that as we have seen host a Yang-Mills
theory, while on the other there is a corresponding 
string background, and J. Maldacena conjectured that the 
resulting string theory 
(which includes gravity) in the bulk of $AdS_5$ 
is exactly {\it equivalent} (dual) to an ${\cal N}=4$ 
${\rm U}(n)$ super Yang-Mills
theory in its border, the four dimensional 
Minkowski space.  This remarkable correspondence actually
reflects 
a number of unusual equivalences between string amplitudes: for instance, 
as shown in figure 13, a one-loop diagram for open strings, obtained 
widening an ordinary field 
theory loop into an annulus, can alternatively 
be regarded as a tree-level diagram for closed strings. In other words, the
distinction between closed and open strings, and thus between gravity and 
gauge fields, is somewhat blurred in String Theory.
The conjecture was particularly well tested in the regime where
the size of the strings is very small compared to the radii of 
$AdS_5$ and $S^5$ and where the string coupling constant is also small,
so that the string theory is well described by classical supergravity.
In the dual picture, this corresponds to the ${\rm U}(n)$ Yang-Mills theory in 
the limit where both $n$ and the 't Hooft coupling $g^2_{YM}n$ are large, 
{\it i.e.} in its deep quantum mechanical regime. Still, some quantities
protected by supersymmetry match mirably in the two descriptions, 
confirming this surprising 
correspondence between 
theories defined in different space-time dimensions.
Tests at intermediate regimes are much harder and are still largely lacking,
but no contradictions have emerged so far. Gravity would 
this way ideally provide a tool to study quark confinement, but with a
new ingredient: the color flux tubes penetrate an additional dimension
of space time.

We have thus come to a full circle somehow. The string idea originated from 
attempts made in the sixties to model the strong interaction amongst mesons via
narrow flux tubes, that culminated in the famous work of G. Veneziano. 
With the advent of QCD, this picture was abandoned, since
the flux tubes were regarded as a manifestation of QCD itself, while 
strings were proposed, as we have seen, as a tool to attain a finite 
quantum gravity. However, many people kept looking for a string-like
description of the color flux-tubes, and with the advent of the AdS/CFT 
correspondence this was indeed realized to some extent, albeit once more
in a supersymmetric setting that is free of many intricacies of QCD. 
Again, difficulties of various types are met when one tries to 
proceed away from supersymmetry to come closer to our real confining 
low-energy world.

\section*{Acknowledgments}

We are grateful to the Organizers for their kind invitation to the 
2002 Moriond Electro-Weak Meeting, and for their 
encouragement to merge our contributions in this short review aimed 
mostly at experimental particle physicists.
The work of the first author was supported in part by I.N.F.N., by the 
European Commission RTN programmes HPRN-CT-2000-00122 and 
HPRN-CT-2000-00148, by the INTAS contract 
99-1-590, by the MURST-COFIN contract 2001-025492 and by the NATO
contract PST.CLG.978785.
The work of the second author was supported in part by the ``FWO-Vlaanderen''
through project G.0034.02, by the Federal Office for Scientific, Technical and
Cultural Affairs through the Interuniversity Attraction Pole P5/27 and by the 
European Commission RTN programme HPRN-CT-2000-00131, in which he
is associated to the University of Leuven.

We are grateful to C. Angelantonj, J. Lemonne, J. Troost, G. Stefanucci and
F. Zwirner for useful suggestions and comments on the manuscript.
\vskip 15pt
The spirit of this review suggests that we refrain from giving detailed
references to the original literature, contenting ourselves with a number 
of books and reviews that introduce the various topics addressed in this 
paper, and where the interested reader can find further details.

\end{document}